\documentclass[conference]{IEEEtran}

\IEEEoverridecommandlockouts

\usepackage{cite}
\usepackage{amsmath,amssymb,amsfonts}
\usepackage{algorithmic}
\usepackage{graphicx}
\usepackage{textcomp}
\usepackage{xcolor}
\usepackage{makecell}
\usepackage{caption}
\usepackage{makecell}
\usepackage{multirow}
\usepackage{siunitx}

\usepackage[
    colorlinks=true,
    urlcolor=blue,
    linkcolor=blue,
    citecolor=blue
]{hyperref}

\newcolumntype{C}{>{\centering\arraybackslash}X}

\def\BibTeX{{\rm B\kern-.05em{\sc i\kern-.025em b}\kern-.08em
    T\kern-.1667em\lower.7ex\hbox{E}\kern-.125emX}}
\DeclareMathAlphabet\mathbfcal{OMS}{cmsy}{b}{n}

\begin{document}

\title{\LARGE \bf
A Programmable Optics Cloud Laboratory
}

\author{Sachin Vaidya$^{1,2\bigstar}$, Caio Silva$^{1\bigstar}$, Seou Choi$^{1\bigstar}$, Joshua Chen$^{1}$ and Marin~Solja\v{c}i\'{c}$^{1,2}$
\thanks{$^\bigstar$ These authors contributed equally. Correspondence should be directed to: \tt\small\{svaidya1, caiosiq, seouc130\}@mit.edu}
\thanks{$^{1}$ Sachin Vaidya, Caio Silva, Seou Choi, Joshua Chen and Marin~Solja\v{c}i\'{c} are with Research Laboratory of Electronics, Massachusetts Institute of Technology, Cambridge, MA 02139, USA}
\thanks{$^{2}$ Sachin Vaidya  and Marin~Solja\v{c}i\'{c} are with NSF Institute for Artificial Intelligence and Fundamental Interactions and the Department of Physics, Massachusetts Institute of Technology, Cambridge, MA 02139, USA}}

\maketitle

\begin{abstract}
Laboratory automation can improve experimental throughput, accessibility, and reproducibility, but many robotic laboratory systems remain difficult to reconfigure. This challenge is especially pronounced in free-space optics, where experiments are built from heterogeneous components, require precise alignment, and are frequently rearranged as experimental goals change. In this work, we present the Programmable Infrastructure for Cloud Optics (PICO), a robotic cloud-laboratory architecture designed to make reconfigurable optical experiments easier to program, operate, and reproduce. PICO provides a common domain-specific abstraction and software layer through which experimental configurations and actions can be controlled across different user interfaces. This enables the same physical laboratory to support remote interactive use, scripted experiments, autonomous routines, and features such as version control. We implement PICO on a robotic free-space optics platform and demonstrate it through an experimental case study. \href{https://anonymous.4open.science/r/_PICO_}{Project Link.}

\end{abstract}
\vspace{-6 pt}
\section{Introduction}

Laboratory automation is transforming experimental science by improving throughput, reducing repetitive manual work, and enabling closed-loop experimentation. In chemistry, materials science, and biology, robotic platforms have already demonstrated automated synthesis, characterization, and optimization workflows~\cite{zhang2025multimodal, darvish2025organa, rapp2024self}. Cloud laboratories extend this idea by connecting experimental infrastructure through software interfaces, allowing users to operate physical instruments remotely. A central requirement for such systems is therefore not only robotic automation, but also a general representation through which experiments can be specified and executed.

Free-space optics provides a particularly challenging and important domain for this problem. Optical experiments are used throughout imaging, spectroscopy, metrology, quantum information,  photonics, and many other areas of science and engineering. A free-space optical experiment is generally constructed directly on an optical table from a heterogeneous collection of mirrors, lenses, polarizers, wave plates, light sources, cameras, and detectors. Building and operating such setups is time-consuming and expertise-intensive, often taking months of effort: components must be placed, iteratively aligned, and frequently re-adjusted as the experiment is modified. The positions and orientations of components often need to be controlled with sub-millimeter or sub-degree precision and small perturbations can render entire optical setups completely unusable. This combination of reconfigurability, hardware heterogeneity, and precision demands has kept much of experimental optics dependent on manual operation.

Recent work has shown that many of the physical challenges of automating optics can be overcome. Robotic platforms can now assemble optical setups, perform precision alignment, execute measurements, and recover from disturbances~\cite{uddin2026aidriven,choi2026closedloop}. These capabilities are implemented as task-specific workflows rather than as operations on a common representation of the components and the laboratory. Such a representation is particularly important in optics, where the same hardware can be reconfigured into fundamentally different experiments. A shared abstraction would also allow optical experiments to be treated as programmable objects: configurations could be saved, compared, restored, and executed through common interfaces rather than being encoded in hardware-specific scripts. Beyond improving reproducibility, this could enable remote access to sophisticated optical experiments, facilitate sharing of experimental procedures across laboratories, and provide a foundation for autonomous experimental science.

\begin{figure}[t]
\centering
\includegraphics[width=\columnwidth]{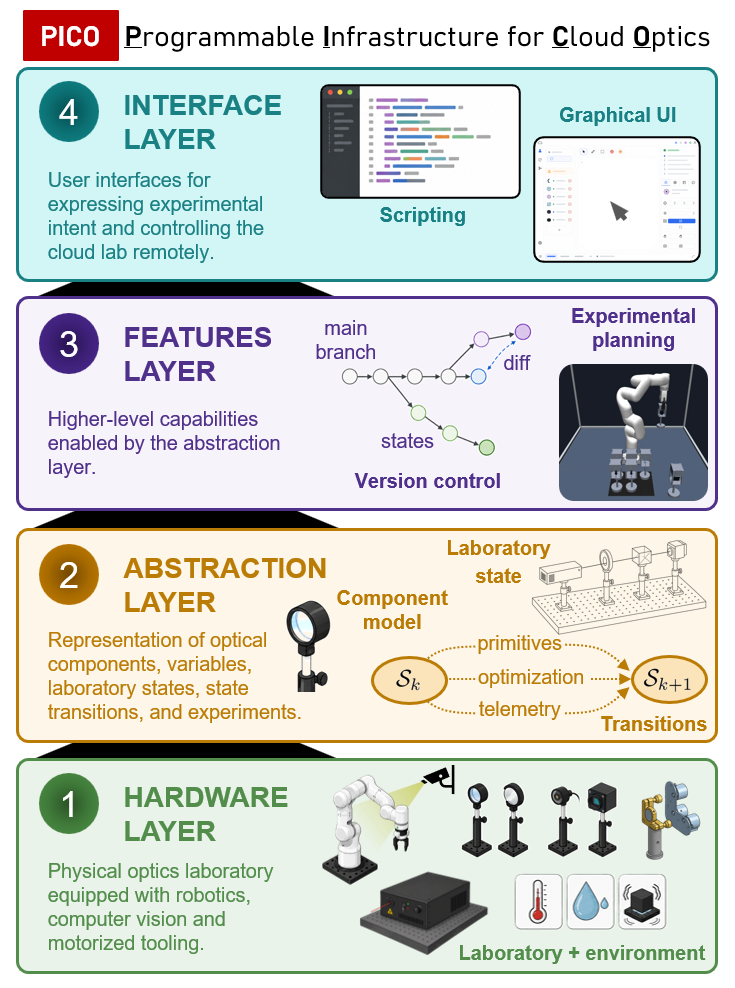}
\caption{Overview of PICO, organized into four layers: the physical hardware; the laboratory abstraction of components, state, and state transitions; higher-level features enabled by the abstraction; and the graphical and scripting interfaces through which users can control the laboratory remotely.}
\label{fig:PICO}
\vspace{-12pt}
\end{figure}

In this work, we present the first programmable optics cloud laboratory for reconfigurable free-space optics. We call the underlying architecture \textbf{PICO}, which stands for \underline{P}rogrammable \underline{I}nfrastructure for \underline{C}loud \underline{O}ptics (summarized in Fig.~\ref{fig:PICO}). The laboratory combines general-purpose robotic manipulation and sensing with a domain-level abstraction that formalizes what laboratory components are, which of their properties can be controlled, what can be measured from the experiment, and how robotic operations transform the laboratory state. This provides scientists with a unified representation across user interfaces, scripts, teleoperation, and autonomous routines, decoupling experimental intent from the hardware. As a result, physical experiments become more amenable to the same concepts that make software programmable and reproducible, including version control, validation and debugging, and reusable procedures. We implement PICO on a robotic free-space optics platform and demonstrate these capabilities through the measurement of the spin Hall effect of light. This experimental workflow provides a representative optics example in which components must be exchanged and realigned while preserving a shared experimental configuration.

\section{Related Work}
Recent work has established that free-space optical experiments can be automated with high precision. General-purpose robotic platforms have integrated computer vision, robotic manipulation, and AI-based design to assemble, align, and characterize reconfigurable optical systems~\cite{uddin2026aidriven,choi2026closedloop}. Other studies have focused on narrower alignment problems, including reinforcement-learning-based alignment of optical interferometers~\cite{sorokin2020interferobot,makarenko2022aligning}. Together, these works establish many of the perception, manipulation, measurement, and closed-loop control capabilities required for optics, but these are primarily focused on particular experimental workflows or robotic tasks rather than a common representation of the laboratory.

More general abstractions have been developed in chemistry and the life sciences. ARChemist represents chemistry workflows using human-readable recipes together with explicit models of samples, stations, operations, and workflow state~\cite{fakhruldeen2022archemist}. Aquarium integrates experimental design, inventory, protocol execution, and data capture within a common laboratory software environment~\cite{vrana2021aquarium}, while LabOP provides a hardware-independent representation of biological protocols that can be specialized to different execution environments~\cite{bartley2023building}. At the orchestration level, ChemOS 2.0 coordinates experimental and computational resources within self-driving laboratories~\cite{sim2024chemos}, while AlabOS provides a reconfigurable workflow-management framework for autonomous laboratories~\cite{alabos}. OSCAR similarly provides a modular robotic architecture for biological laboratories, mapping domain-level operations to reusable robotic behaviors~\cite{pivin2026oscar}.

These systems demonstrate the value of separating experimental intent from particular hardware implementations. Free-space optics, however, poses a distinct challenge: experiments are inherently geometric, with functionality determined by the spatial arrangement of components, while also imposing stringent precision requirements. This work focuses on developing and implementing a programmable abstraction tailored to optics, thus enabling a cloud laboratory in this domain.

\section{Overview of the Hardware Platform}
\label{sec:hardware}

PICO is implemented on a reconfigurable free-space optics platform built around a 7-DOF robotic arm (UFACTORY xArm7) operating on a standard optical table (Fig.~\ref{fig:state_transitions}a). The robot provides the primary means of component handling and placement. To make heterogeneous optical components (such as mirrors, lenses, beam splitters, polarizers, wave plates, and detectors) compatible with robotic manipulation, they are placed in standardized 3D-printed housings and bases. These fixtures provide reliable grasping geometry while preserving the modularity of conventional laboratory components. Components in the workspace are identified and localized using fiducial markers (ArUco) and computer vision. Overhead stereo cameras provide workspace-level component identification and coarse pose estimation, while an end-effector-mounted depth camera refines component pose during grasping and placement. Optical cameras and other detectors provide experiment-specific measurements of the laser beam.

In addition to robotic pick-and-place, the platform supports fine actuation of component degrees of freedom through motorized stages and alignment tools. These include rotational or translational stages and actuated controls on standard optical mounts, enabling fine adjustments to quantities such as mirror angle, polarizer orientation, and wave-plate angle.

The physical laboratory is connected to PICO's software components through an edge computer: a local computer co-located with the experimental platform that provides the software interface between PICO and the laboratory hardware. This edge process directly owns the robotic arm, cameras, motorized stages, and other instrument drivers, and translates abstraction-level commands into hardware-specific operations. Remote user interfaces do not communicate with these devices directly; instead, requests are sent through a central coordinator that performs session management and command validation before routing them to the appropriate laboratory edge.

\section{Architecture and Laboratory Abstraction}
\label{sec:abstraction}

Free-space optics has traditionally lacked an explicit computational representation of experimental states and actions. A programmable cloud laboratory requires a description of the experiment that is independent of the particular interface or execution mode through which it is controlled. Here, we formalize such a sensible shared laboratory abstraction for free-space optics as a core component of PICO.

\subsection{Component Model and Definitions}
\label{sec:components}

A \emph{component} is any individually addressable physical object in the laboratory, such as a mirror, lens, polarizer, wave plate, camera, detector, or laser. Each component indexed by $i$ is associated with a component definition
\begin{equation}
    \mathcal{C}_i =
    \left(
    \mathcal{P}_i,
    \mathcal{T}_i,
    \mathcal{M}_i
    \right),
    \label{eq:component_definition}
\end{equation}
where $\mathcal{P}_i$ denotes its parameters, $\mathcal{T}_i$ the set of declared tunables, and $\mathcal{M}_i$ the set of declared measurables.

\emph{Parameters} encode properties used to describe the identity and capabilities of the component. These are considered fixed and unmodified during experiment execution. Examples include the component name, focal length of a lens, or the type of anti-reflection coating on a mirror.

A \emph{tunable} declaration specifies a directly controllable degree of freedom exposed by the component. Examples include the Cartesian position of a component, the motor positions of a rotation stage, camera exposure time, or a laser power setpoint. The declaration may include its datatype, admissible range, units, and supported control operations. Its value, $\mathbf{t}_{i,k}$, at some instance $k$ is stored and handled separately, as discussed below.

A \emph{measurable} declaration specifies an observable quantity exposed by the component. Examples include a camera image, a photodiode voltage, beam centroid, or beam width. The treatment of its value is also discussed below.

In the implementation, these declarations are maintained in a component catalog, which specifies the parameters, tunables, and measurables, and supported primitives (also described below) associated with each physical component.

\subsection{Laboratory State and Tunables}
\label{sec:state}
We define the laboratory state as the set
\begin{equation}
    \mathcal{S}_k
    =
    \left\{
    \left(
    \mathcal{C}_i,
    \mathbf{t}_{i,k}
    \right)
    \right\}_{i=1}^{N},
    \label{eq:lab_state}
\end{equation}
where $N$ is the number of components currently represented in the laboratory. Thus, the state explicitly associates each component with the current values of its tunables $\mathbf{t}_{i,k}$. Tunables are state variables: their values can be directly addressed by the control system and, within the assumed actuation tolerance, intentionally modified by the user. For example, changing a mirror angle, translating a lens, or modifying a camera exposure setting changes their corresponding tunables and therefore changes the laboratory state. We note that although commanded and physically-realized tunable values may differ in practice, we assume them to be equal within the tolerances of the laboratory, which we have operationally verified through experiments. We also emphasize that the term ``laboratory state" is used here to mean the restorable logical configuration exposed by the abstraction, rather than the complete physical state of the laboratory.

\subsection{Measurables and Measurement Kernels}
The values of measurables represent experimental observations. These values, $\mathbf{m}_{i,k}$, at some instance $k$ generally depend on the configuration of the complete experiment. This can be expressed as:
\begin{equation}
    \mathbf{m}_{i,k}
    =
    h_i\!\left(
    \mathcal{S}_k,
    \mathbf{e}_k,
    \xi_k
    \right),
    \label{eq:observation}
\end{equation}
where $\mathcal{S}_k$ is the laboratory state, $\mathbf{e}_k$ represents environmental and uncontrolled or unmodeled variables, $\boldsymbol{\xi}_k$ represents stochastic measurement uncertainty, which may be quantified, and $h_i$ is a function that depends on these quantities. Measurable values therefore cannot, in general, be directly set or reconstructed from commanded tunable values alone. Consequently, a quantity derived from $\mathbf{m}$ remains a measurable since deterministic post-processing does not remove its dependence on $\mathbf{e}$ or $\xi$. We define such deterministic processing operations as \emph{measurement kernels}. A kernel $\kappa$ maps an acquired measurable to another measurable, $\mathbf{m}'_{i,k}=\kappa(\mathbf{m}_{i,k})$, for example transforming a camera image of a laser beam into a beam centroid, integrated optical power, or beam width. Kernels provide a shared vocabulary for processing measurables and operate only on already acquired data.

\subsection{State Transitions}
\label{sec:primitives}

\subsubsection{Primitives}
We expose physical operations through \emph{primitives}: validated laboratory operations/actions defined over components (Fig.~\ref{fig:state_transitions}b). Some primitives include moving a component to a specified position, rotating a component, and changing hardware setpoints. The physical backend maps a primitive to robot trajectories, motor commands, etc. A primitive that modifies a tunable induces a state transition $\mathcal{S}_{k+1} = T_{\mathrm{primitive}}(\mathcal{S}_k, a_k, \boldsymbol{\eta})$, where $a_k$ identifies the requested primitive and $\boldsymbol{\eta}$ denotes additional execution parameters, such as motion or actuation speed.

Not all primitives modify the laboratory state. Measurement or data acquisition operations leave $\mathcal{S}_k$ unchanged while returning values of parameters, tunables or measurables. Such primitives are identity operations with respect to the laboratory state, although they may perform sensing or computation in the physical backend. This retains both the state-mutating primitives and non-mutating primitives within the same structure.

In the implementation, primitives span robotic pick-and-place operations, storage-area management, motor and instrument setpoints, and scientific data acquisition. Primitive availability is declared per component, such that unsupported operations can be identified and rejected before execution.

\subsubsection{Optimization}
\label{sec:optimization}

\begin{figure}[ht]
\centering
\includegraphics[width=0.5\textwidth]{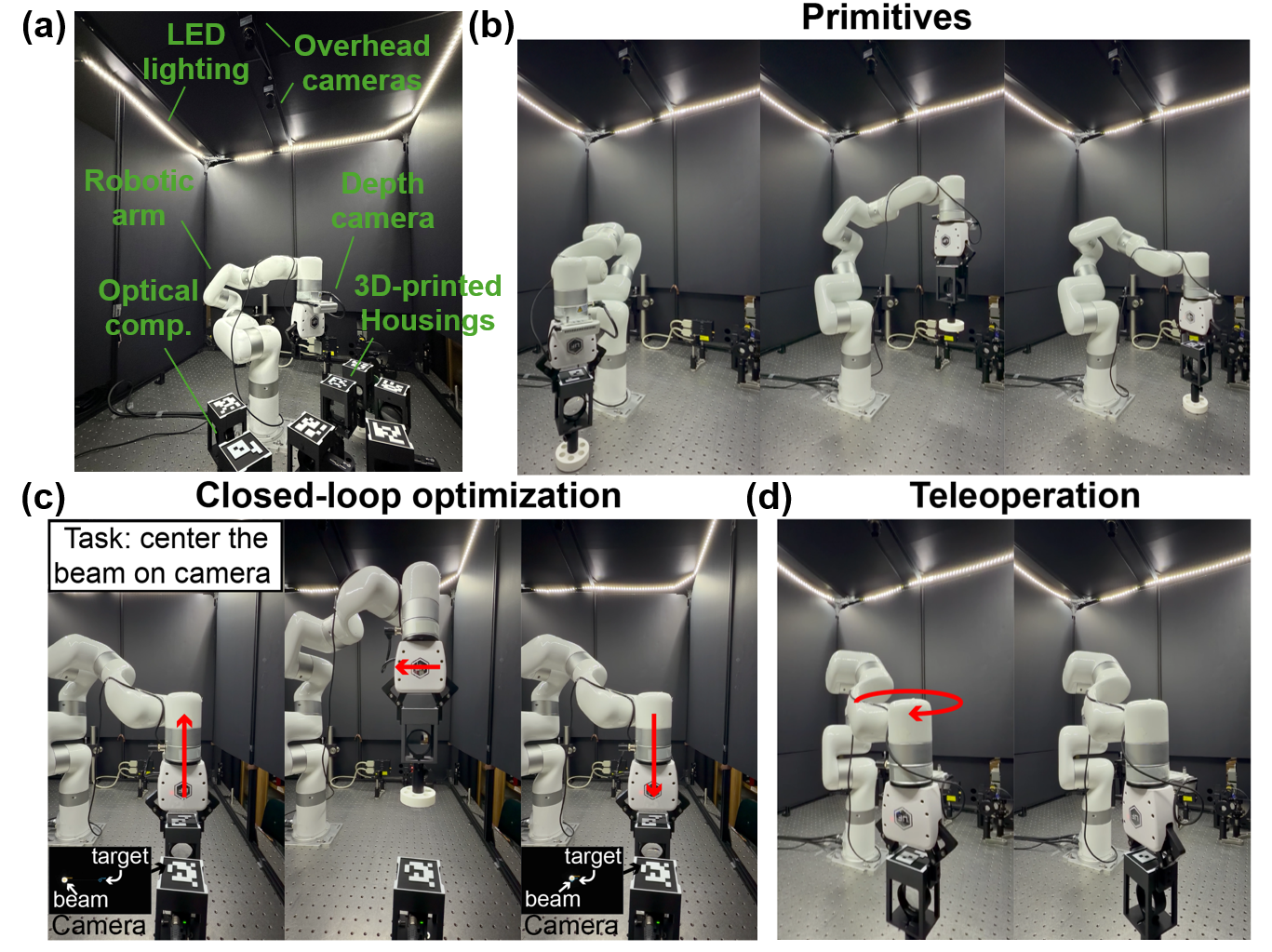}
\caption{(\textbf{a}) Overview of the hardware platform, which includes the xArm7 robotic arm, overhead stereo cameras, end-effector depth camera, standard optical table, workspace illumination, and 3D-printed optical component housings and commercially available motorized mounts. (\textbf{b})-(\textbf{d}) Three types of implemented state transitions: Primitives (showing moving a component), Closed-loop optimization (showing the centering of a beam via positional adjustments to a lens), Teleoperation (showing a continuous rotation of a component by a user).}
\label{fig:state_transitions}
\vspace{-12pt}
\end{figure}

This provides another state-transition mechanism in which the sequence of actions is determined dynamically from measured feedback (Fig.~\ref{fig:state_transitions}c). Let $\tau$ denote the selected set of tunables to be optimized, $\Omega(\mathbf{m})$ denote an objective defined in terms of one or more measurables $\mathbf{m}$, and $\mathcal{A}$ denote the optimization algorithm specification including its hyperparameters and termination criteria. We represent a general optimization routine as $\mathcal{S}_{k+1}
    =
    T_{\mathrm{opt}}
    \left(
    \mathcal{S}_k,
    \tau,
    \Omega(\mathbf{m}),
    \mathcal{A}
    \right).$
    
For example, $\Omega$ may specify a target beam size or mode profile, minimization of a measured error, or maximization of detected optical power. Importantly, such optimization routines can be made component-independent. The same optimization could, e.g., jointly adjust a component's Cartesian position, a motorized mirror angle, and an instrument setting using an objective derived from a camera image. Internally, $T_{\mathrm{opt}}$ may perform an arbitrary number of measurement and actuation cycles and traverse many laboratory states before terminating. These internal states are hidden at the abstraction level, with only the resultant end state recorded as the next state $\mathcal{S}_{k+1}$.

\subsubsection{Teleoperation}
\label{sec:teleoperation}

Some laboratory operations are more naturally performed through interactive user control than as pre-specified commands. We expose these through teleoperation, in which a user continuously manipulates one or more tunables while receiving live sensor feedback (Fig.~\ref{fig:state_transitions}d). At the abstraction level, the detailed sequence of continuous user inputs and physical configurations is encapsulated within the teleoperation process: $\mathcal{S}_{k+1}
    =
    T_{\mathrm{teleop}}
    \left(
    \mathcal{S}_k
    \right).$ Similar to optimization, we represent a teleoperation session only through its initial and final laboratory states.

\subsection{Experimental Trace}
\label{sec:experimental_trace}

We define an \emph{experimental trace} as a state-transition trace given by
\begin{equation}
    \mathcal{L}:
    \quad
    \mathcal{S}_1
    \xrightarrow{T_1}
    \mathcal{S}_2
    \xrightarrow{T_2}
    \mathcal{S}_3
    \xrightarrow{T_3}
    \mathcal{S}_4
    \cdots
    \xrightarrow{T_{K-1}}
    \mathcal{S}_K,
    \label{eq:trace}
\end{equation}
where each transition $T_j$ is chosen from the available transition mechanisms. Thus, a trace records the sequence of abstraction-level laboratory states and the transition mechanism used to move between them; internal states generated by $T_{\mathrm{teleop}}, T_{\mathrm{opt}}$ are not included. Non-mutating primitives, such as for data acquisition operations, may also appear in a trace. These correspond to identity transitions from the perspective of the state. These can be represented in the trace as $\cdots \xrightarrow{} \mathcal{S}_k\xrightarrow{T_m}\mathcal{S}_k \xrightarrow{} \cdots$, where $T_m$ is some non-mutating primitive that performs a measurement and acquires a measurable.

An experiment is therefore not, in general, specified by only its final state $\mathcal{S}_K$. Many experiments consist of acquiring data across a sequence of distinct configurations. Two executions may reach the same final laboratory state while producing different experimental data because they followed different intermediate states or acquired different measurements. The trace therefore preserves the abstraction-level history needed to contextualize the measurements produced during execution.

\subsection{Experimental Recipe}
\label{sec:experimental_recipe}
An experimental recipe is an executable program or procedure expressed in terms of the components, tunables, measurables, and state-transition mechanisms defined above. Recipes may include conventional programming constructs such as conditionals, loops, and functions. These general-purpose programming constructs are supplied by the host programming environment and are kept separate from the laboratory abstraction. Given a compatible laboratory implementing the same abstraction and capabilities, a recipe specifies how the experiment can be re-executed on another laboratory instance. Executing a recipe produces an experimental trace together with the measurements acquired along that trace.

\section{Capabilities of PICO}
\label{sec:capabilities}

The abstraction of components, laboratory state, and actions enables capabilities beyond basic hardware control. We highlight four examples of such capabilities of PICO in subsections~\ref{sec:remote_interfaces} - \ref{sec:digital_twin} below.

\subsection{Unified Remote Interfaces}
\label{sec:remote_interfaces}

The laboratory can be accessed and operated remotely through both graphical and scripting user interfaces (UIs), providing the basis for its use as a cloud laboratory (Fig.~\ref{fig:capabilities}a). In either case, users interact with components through their tunables, measurables, and state transitions rather than through low-level hardware or robot commands. This common representation allows interactive and programmatic workflows to remain compatible within the same architecture. A user may, for example, explore or align a setup through the graphical interface and then encode the resulting procedure as a reproducible recipe, while changes made through either interface remain reflected in the same laboratory state.

PICO is implemented as a client–coordinator–edge architecture. The client layer consists of the browser-based graphical user interface and a Python SDK, both of which issue the same abstraction-level primitive commands. A central coordinator validates primitive requests against component capabilities, manages laboratory state and long-running jobs, and enforces an exclusive session lease so that only one authoring client can mutate a physical backend via the edge at a time. For latency-sensitive closed-loop operations such as optimization, the selected tunables, optimization specification, and required kernels can be transferred to the laboratory edge once, allowing the capture–evaluate–actuate loop to execute locally rather than requiring a network round trip for every iteration. Only progress updates and the resulting abstraction-level state need to traverse the network.

\subsection{Validation and Debugging of Physical Operations}
\label{sec:validation}

A structured representation of components and actions also allows experimental requests to be validated before they are executed physically. Each component declares the tunables and primitives that it supports, together with constraints such as admissible ranges where applicable. A requested primitive can therefore be checked against the corresponding definition before being dispatched to the physical backend.

Validation additionally occurs at the robotic execution layer. Motions can be rejected when they require an unreachable configuration, produce a collision, or violate workspace constraints. Since the failed request remains associated with a specific component and primitive, the system can return errors at the level at which experimental intent was expressed rather than exposing failures only through low-level robot trajectories or device commands. For example, if a requested component motion would intersect another previously placed component, the potential collision can be detected before execution and the system can report exactly which component obstructs the requested path or placement. This provides a form of \emph{physical compilability} for experimental procedures. An experimental program is not merely executable software: its operations are checked against the capabilities and physical constraints of the laboratory on which it is to be executed.

\subsection{Version Control of Laboratory Configurations}
\label{sec:version_control}

Because the laboratory state $\mathcal{S}_k$ explicitly associates each component with the current values of its controllable tunables, experimental configurations can be treated as versionable physical states. Saved laboratory states can be tracked using a Git-like history, allowing users to selectively commit states that are experimentally meaningful and associate them with messages or metadata. Branching provides a natural mechanism for exploring alternative experimental configurations from a common starting point while retaining the ability to return to earlier states. This is useful, for example, when a user reaches a well-aligned optical setup, identifies a useful intermediate stage of an experiment, or wishes to preserve a working configuration before making modifications. The resulting commit history therefore consists of selected states elevated to persistent checkpoints.

A central requirement for such physical version control is that a committed state can actually be restored. This is not generally possible for arbitrary physical operations, for example, unmixing two liquids. By contrast, in the operating regime for optics considered here, the state is defined in terms of tunables, such as component positions, motor angles, and setpoints. Restoring a saved state then reduces to driving these tunables back to their committed values. Note that this operation does not restore measurable values, which must be re-acquired, e.g., by re-optimizing the corresponding tunables.

The restorability of a state is supported by useful algebraic structure in the logical primitive state-transition functions introduced in Sec.~\ref{sec:primitives}. For a relative primitive $T_{\delta}$ acting on a tunable, successive updates compose according to their net change, $T_{\delta_2}\circ T_{\delta_1}=T_{\delta_1+\delta_2}$, and, when the inverse operation is admissible, $T_{-\delta}\circ T_{\delta}=I\text{ (identity)}$. Absolute-setpoint primitives obey overwrite rules: if $T_v$ sets a tunable to the value $v$, then $T_{v_2}\circ T_{v_1}=T_{v_2}$, since only the final commanded value is relevant. Additionally, many logical primitives acting on the tunables of independent components commute. To give a few examples, several relative motor adjustments may be replaced by one net rotation, inverse position adjustments may cancel, and multiple absolute setpoints reduce to the final one. These properties allow restoration to target a state without replaying the state-transition history that originally produced it.

This structure also enables an experimentally meaningful notion of \emph{physical diff}. For two previously saved states, $\mathcal{S}_a$ and $\mathcal{S}_b$, the physical diff, $\Delta(\mathcal{S}_a, \mathcal{S}_b)$, identifies the tunables that must change to move between two committed configurations. These changes can then be realized through the corresponding primitive state-transition functions allowing the laboratory to move from $\mathcal{S}_a$ to $\mathcal{S}_b$ without replaying the original sequence of primitives, teleoperation, or optimization steps that produced either state. The diff specifies a semantic change in configuration; the physical backend remains responsible for realizing it as valid hardware actions.

In the implementation, PICO realizes this version-control functionality using Git-like graph structures: committed laboratory states are stored as immutable snapshots linked by parent relationships, while named branches reference selected commits and the live laboratory configuration acts as the working state. When a user checks out a previous commit, PICO computes the physical diff between the current and target configurations and compiles the required changes into a dependency-aware set of primitive operations. Spatial dependencies are introduced when one component must vacate a region before another can occupy its target location; when cyclic rearrangements occur, temporary staging locations, such as in the storage, can be used to clear the required space before components are moved to their final targets.

\begin{figure}[t]
\centering
\includegraphics[width=0.45\textwidth]{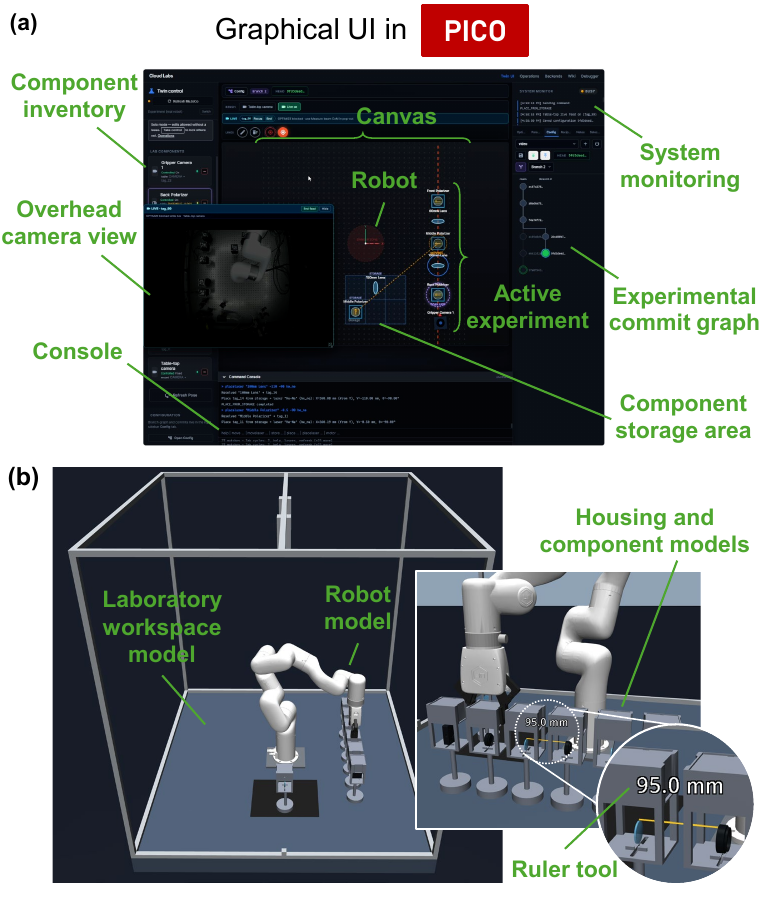}
\caption{(\textbf{a}) Graphical UI for controlling and monitoring the optical cloud laboratory showing implemented capabilities and features. (\textbf{b}) Digital twin of the laboratory, including models of the workspace, robot, optical components and housings, and a ruler tool (MuJoCo). The digital twin is used to visualize and test experimental layouts before physical execution.}
\label{fig:capabilities}
\vspace{-12pt}
\end{figure}

\subsection{Digital Twin for Experimental Layout Planning}
\label{sec:digital_twin}

The same abstraction can also be executed by a virtual backend for experimental planning. As part of PICO, we implement a robotic digital twin of the laboratory, including the robotic arm, optical components, and their standardized housings and bases. The digital twin is controlled through the same UIs as the physical laboratory, allowing user commands to be visualized and executed first in simulation.

This is particularly useful for free-space optics, where experimental layouts are strongly geometric and quantities such as component position, orientation, spacing, and beam-path clearance determine whether a setup can be realized. Users can construct candidate layouts in the virtual environment and inspect the corresponding robotic motions before requesting physical execution. The digital twin need not reproduce every primitive, tunable, or measurable available on the physical platform; only those aspects relevant to layout, robotic feasibility, or visualization need to be represented. In this sense, it serves a role analogous to sketching an optical setup before assembly, but with the geometry expressed directly in the coordinates and hardware constraints of the robotic platform.

We implement the virtual backend in MuJoCo~\cite{todorov2012mujoco} using a model of the xArm7 manipulator together with rigid-body models of the optical components, housings, bases, gripper, and surrounding workspace (Fig.~\ref{fig:capabilities}b). The simulated scene is instantiated from the laboratory state, and interface commands update the corresponding virtual tunables. Pick-and-place operations are simulated using rigid-body dynamics, including component mass and inertia, gravity, contact, and gripper friction. A ruler tool is also available for users to measure distances. A planned configuration is represented by a virtual laboratory state
$\mathcal{S}^{\mathrm{sim}} = \left\{ \left( \mathcal{C}_i, \mathbf{t}^{\mathrm{sim}}_i \right) \right\}_{i=1}^{N}$.
Tunables shared by the virtual and physical backends can then be transferred directly as targets for physical execution, with the physical backend remaining responsible for validation and hardware actuation. Because both backends use the same state representation, virtual configurations can also be compared, versioned, revised, and used as experimental checkpoints before deployment.

\begin{figure*}[!ht]
\centering
\includegraphics[width=0.95\textwidth]{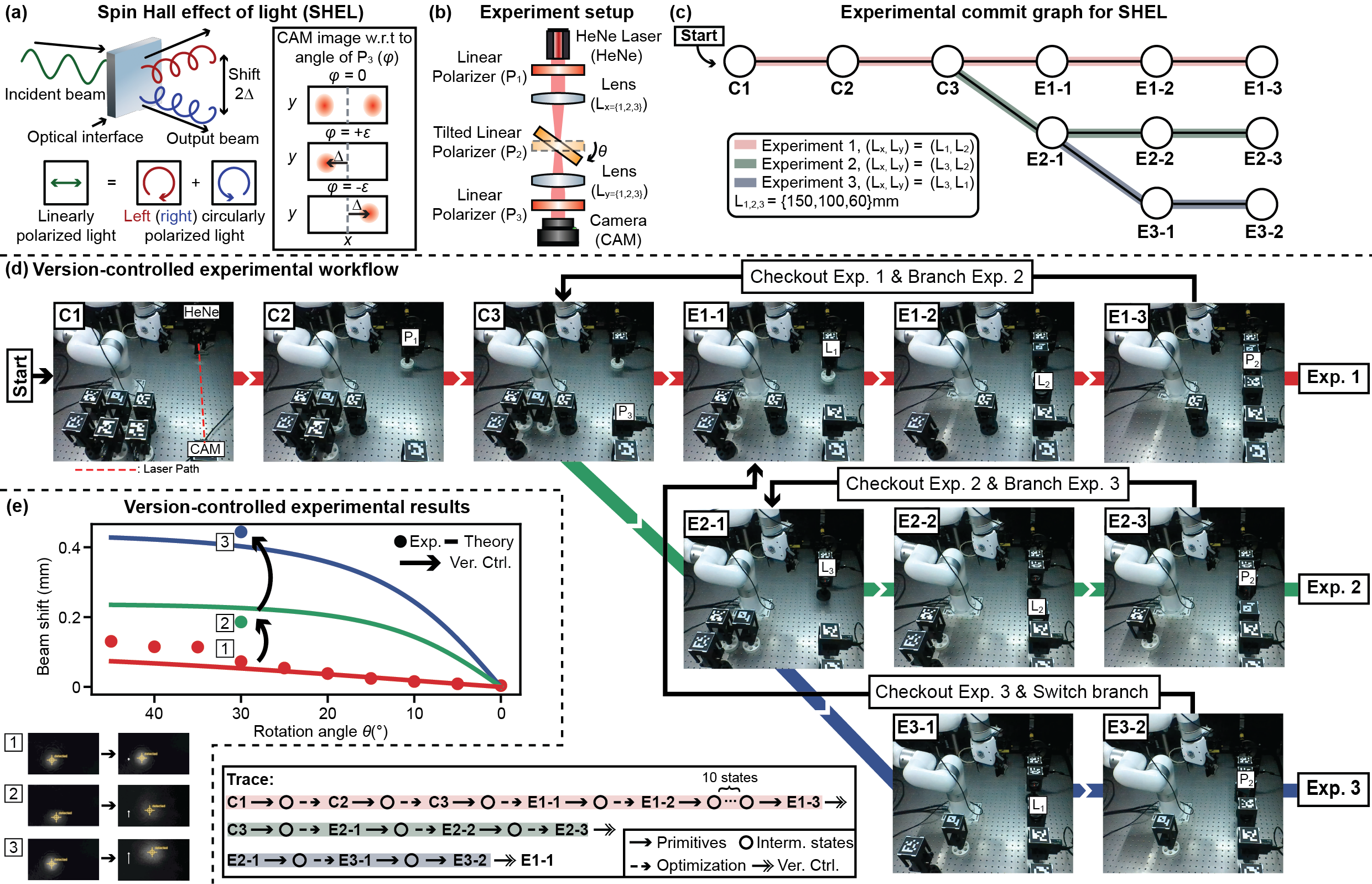}
\caption{Optics experiments conducted using PICO. \textbf{(a)} Schematic of the spin Hall effect of light (SHEL). \textbf{(b)} Schematic of the experimental setup. The lenses are varied throughout the different experiment configurations. \textbf{(c)} Experimental commit graph for SHEL. \textbf{(d)} Version-controlled experimental workflow. The lower panel shows the experimental trace for the full run, along with the state transition functions employed. \textbf{(e)} Experimental results for the various configurations explored.}
\label{fig:SHEL_result}
\vspace{-12pt}
\end{figure*}

\section{Experimental Case Study}

We now present a case study that demonstrates PICO on multiple optics experiments involving the spin Hall effect of light (SHEL) (Fig.~\ref{fig:SHEL_result}a). The SHEL produces beam shifts that depend on the polarization, or optical spin, of light. Their strong sensitivity to small changes at optical interfaces makes them useful for characterizing ultrathin materials and sensing changes in chemical concentration or environmental conditions in real time~\cite{kim2023spin}. Fig.~\ref{fig:SHEL_result}b presents a schematic of the experimental setup used to characterize the SHEL upon transmission through a tilted polarizer. The first lens, $\mathrm{L}_{\mathrm{x}}$, tightly focuses the incident laser beam onto the tilted polarizer, thereby controlling the system's sensitivity to the small SHEL-induced angular shifts associated with different polarization states. The second lens, $\mathrm{L}_{\mathrm{y}}$, converts the angular shift into a measurable spatial displacement at the beam detection camera. Consequently, the measured beam displacement depends on the combination of the focal lengths of the two lenses. 

Fig.~\ref{fig:SHEL_result}c visualizes the experimental commit graph generated using PICO to characterize SHEL-induced beam shifts under different experimental configurations, specifically different combinations of lenses. Three lens configurations are evaluated, each highlighted with a different color. In this case, version control allows each trial to branch from a common state. This approach eliminates the need to reconstruct the shared portion of the setup for every configuration. 

Fig.~\ref{fig:SHEL_result}d illustrates the version-controlled experimental workflow, with a detailed experimental trace shown at the bottom. The experiment begins at state C1, in which the HeNe laser and beam detection camera are placed in the active experimental region, while all other components remain in the storage area at the lower-left corner of the optical table. The subsequent states are C2 and C3, in which linear polarizers P1 and P2 are placed in the laser path using primitives, with their polarization axes set parallel (vertical) to the polarization of the input laser beam using optimization. 

The first experiment follows the red branch and uses lenses expected to produce the smallest SHEL-induced beam shift. The full layout is first planned using the digital twin as shown in Fig.~\ref{fig:capabilities}b. The two lenses are placed sequentially using the primitives and optimization, and the resulting configurations are saved as states E1-1 and E1-2. Because the magnitude of the beam shift depends on the angle $\theta$ of the tilted polarizer $\mathrm{P}_{\mathrm{2}}$, measurements are performed at ten angles ranging from $0^{\circ}$ to $45^{\circ}$ in $5^{\circ}$ increments. The resulting configuration is then saved as state E1-3, along with measurements. 

After completing the first experiment, the system checks out and restores state C3, which corresponds to the common configuration immediately before lens placement, and branches to the second experiment along the green path. To increase the SHEL-induced beam shift, the first lens is replaced with one having a shorter focal length, which focuses the beam more tightly at the tilted polarizer. Following a workflow similar to that used in the first experiment, the two lenses are placed sequentially, producing states E2-1 and E2-2. The SHEL-induced beam shift is then measured at $\theta = 30^{\circ}$ in E2-3 and compared with that obtained using the first lens configuration.

The focal length of the second lens can be further optimized based on the distance between the beam detection camera and the incident laser to produce a larger spatial beam displacement. We therefore check out and restore state E2-1 before creating a third experimental branch. Because the first lens is already present in state E2-1, only the second lens is replaced with a lens different from that used in the second experiment, resulting in state E3-1. The SHEL-induced beam shift is again measured at $\theta = 30^{\circ}$, after which the configuration is saved as state E3-2. Finally, to evaluate the system's ability to restore a substantially different configuration in the commit graph, the experiment moves from state E3-2 to state E1-1 using the version-control functionality. This demonstrates that the system can reproducibly recover a previously committed experimental state across distinct branches of the workflow. 

Fig.~\ref{fig:SHEL_result}e summarizes the SHEL-induced beam shifts measured under the various experimental configurations explored using version control. The experimental measurements agree well with the theoretical curves where we find that at a fixed $\theta$, the magnitude of the beam shift increases as the lens configuration is varied. To characterize experimental repeatability, we performed two additional measurement runs using the configuration of Experiment 1, which produced the smallest beam shifts, yielding three trials at each of seven angles from $\theta = 0^{\circ}$ to $30^{\circ}$. Across these trials, the overall root-mean-square deviation (RMSD) about the angle-wise means was \SI{3.87}{\micro\meter} (9.7\% of their RMS), or 1.61 camera pixels at a pixel pitch of \SI{2.4}{\micro\meter}. This sub-two-pixel RMS variation indicates good measurement precision and provides a conservative upper bound on repeatability error.

To evaluate the benefit of using version control to explore different experimental configurations, we conducted five end-to-end experiments, each involving three version-control operations, for a total of 15 restoration attempts (Table~\ref{tab:statistics}). Across these runs, we evaluated two distinct commit histories for the same optical experiment and component set. The system successfully completed all five experiments and all 15 restoration operations without execution failure or manual intervention. Table~\ref{tab:statistics} also reports the time required to restore a target state through version control compared with rebuilding the same configuration from the initial state. The restoration cases are grouped according to a graph distance ratio, $\rho$, with five attempts per group, which represents the restoration distance relative to the distance required to construct the target configuration from the initial state. For example, $\rho=2/3$ for the restoration from E2-3 to E2-1 in Fig.~\ref{fig:SHEL_result}c. For $\rho=2/3$, rebuilding time was measured using an executed transition to a configuration differing from the exact target only in one lens's focal length, with negligible time impact, as the exact transition did not naturally occur in the experimental workflow. The results show greater time savings when users modify components introduced near the end of the experimental pipeline, as commonly occurs in optical experiments such as microscopy and material characterization, which require frequent changes of samples, lenses, and filters. 


\captionsetup[table]{labelsep=colon, skip=0pt} 
\begin{table}[t]
\centering
\caption{State-restoration performance}
\label{tab:statistics}
{\setlength{\tabcolsep}{3pt}
\begin{tabular}{|c|c|c|c|c|}
\hline
\multirowcell{2}{Completed\\experiments} &
\multirowcell{2}{Successful\\restorations} &
\multicolumn{3}{c|}{Time required (\% of rebuilding from start)} \\
\cline{3-5}
& & $\rho=2/3$ & $\rho=4/3$ & $\rho=3/2$ \\
\hline \hline
5/5 (100\%) &
15/15 (100\%) &
$11.9 \pm 1.4$ &
$20.6 \pm 3.1$ &
$27.8 \pm 2.0$ \\
\hline
\end{tabular}
}
\vspace{-12pt}
\end{table}

\section{Discussion and Conclusion}

In this work, we introduced PICO, a robotic cloud-laboratory architecture for reconfigurable free-space optics. By representing optical components and laboratory operations through a common abstraction, this architecture allows researchers to remotely manipulate optics experiments through graphical interfaces, scripts, teleoperation, optimization routines, and version-control workflows. We implemented this architecture on a robotic free-space optics platform and demonstrated it through a detailed experimental case study.

Several directions could extend the present system. The MuJoCo digital twin could incorporate optical ray tracing and custom robot motion routines, allowing both optical behavior and robotic feasibility to be evaluated before physical execution. PICO could also be extended to coordinate multiple robots and laboratory workcells through the same abstraction. Another important direction is integration with agentic AI. Rather than requiring an agent to generate low-level robot and instrument control code, the abstraction exposes a constrained, high-level vocabulary. This could allow advances in code generation and scientific agents to transfer more naturally to physical experimentation while leaving hardware-specific execution to the laboratory backend. Finally, deploying PICO as a cloud laboratory for external users will be important for evaluating usability, reproducibility, and the extent to which experimental procedures can be shared and transferred across users and, ultimately, across different laboratory implementations.

\section*{Acknowledgment}

The authors thank Shiekh Zia Uddin, Sahil Pontula, Ryan Lopez, Serena Landers, Jehyeon Shin, Debadarshini Mishra, and Hae Won Lee for stimulating discussions throughout this project. S.C. acknowledges support from the Korea Foundation for Advanced Studies Overseas PhD Scholarship. J.C. acknowledges support from the NSF GRFP under Grant Number 2141064. S.V. and M.S. acknowledge support from NSF under Cooperative Agreement PHY-2019786 (The NSF AI Institute for Artificial Intelligence and Fundamental Interactions). C.S. acknowledges the support of the MIT Undergraduate Research Opportunities Program (UROP). This work was also supported in part by the U.S. Army DEVCOM ARL Army Research Office through the MIT Institute for Soldier Nanotechnologies under Cooperative Agreement W911NF-23-2-0121, the MIT Generative AI Impact Consortium (MGAIC), and Shell International Exploration and Production Inc.

\textbf{\textit{Author Contributions}}— S.V. conceived the project and developed the laboratory abstraction and capabilities with input from S.C. C.S. implemented the abstraction, capabilities, and user interfaces with input from S.V. J.C. implemented the digital twin. S.C. performed experiments and collected the data with input from S.V. M.S. and S.V. supervised the project. All authors contributed to the writing of this manuscript.

\textbf{\textit{{Generative AI Use Disclosure}}}— Generative AI models and tools were used to assist with manuscript copyediting (GPT-5.6 Sol), to generate selected graphical elements in Fig. 1 (Gemini 3.1 Pro), narrate the supplementary video (GPT-5.6 Sol), and to assist with the generation and editing of code for the cloud-laboratory user interfaces and backend (Codex, Cursor Composer 2.5). The authors take responsibility for the final content of this work, including all text, claims, software, figures, and videos.


\end{document}